\documentclass[11pt]{article}

\usepackage[top=1in, bottom=1in, left=1in, right=1in]{geometry}
\usepackage[nodisplayskipstretch]{setspace}
\setdisplayskipstretch{0.3}
\usepackage[utf8]{inputenc}
\usepackage{amsmath}
\usepackage{amsfonts}
\usepackage{bm}
\usepackage{dsfont}
\usepackage{graphicx}
\usepackage{nicefrac}

\usepackage{natbib}
\setcitestyle{authoryear, aysep={}, yysep={;}}

\usepackage[]{endfloat}

\usepackage[shortlabels]{enumitem}
\setitemize{noitemsep,topsep=10pt,parsep=0pt,partopsep=0pt}
\setenumerate{noitemsep,topsep=10pt,parsep=0pt,partopsep=0pt}

\usepackage{authblk}

\usepackage{textalpha}

\usepackage{multirow}
\usepackage[table]{xcolor}
\usepackage{booktabs}
\usepackage[labelfont=bf]{caption}

\usepackage[subtle, mathspacing=normal]{savetrees}
\usepackage[compact]{titlesec}

\graphicspath{ {./paper_figures/} }

\title{Dynamic borrowing from historical controls via the synthetic prior with covariates in randomized clinical trials}

\author[1, *]{Daniel E. Schwartz}
\author[2]{Yuan Ji}
\author[3]{Li Wang}

\affil[1]{Massachusetts General Hospital \& Harvard Medical School, Boston, MA}
\affil[*]{deschwartz.stat@gmail.com}
\affil[2]{Department of Public Health Sciences, \protect\\ 
University of Chicago, Chicago, IL}
\affil[3]{AbbVie, Lake Bluff, IL}

\date{}

\begin{document}

%TC:ignore
\maketitle

\begin{abstract}
  \noindent  Motivated by a rheumatoid arthritis clinical trial, we propose a new Bayesian method  called SPx, standing for synthetic prior with covariates, to borrow information from historical trials to reduce the control group size in a new trial. The method involves a novel use of Bayesian model averaging to balance between multiple possible relationships between the historical and new trial data, allowing the historical data to be dynamically trusted or discounted as appropriate. We require only trial-level summary statistics, which are available more often than patient-level data. Through simulations and an  application to the rheumatoid arthritis trial we show that SPx can substantially reduce the control group size while maintaining Frequentist properties.
\end{abstract}


%TC:endignore

\newpage

\section{Introduction}
There is an explosive interest in utilizing historical control data to improve the design and analysis of a future trial, both in terms of methodological research and clinical trial conduct (Viele et al. 2014). As a motivating application, we consider the development of a next-generation drug in rheumatoid arthritis (RA). Adalimumab has been a standard of care for many patients with RA since its approval in 2003,  consequently serving as the control arm in trials of new RA drugs. 
Since adalimumab has been tested in many trials during its own development as well as in studies of other drugs, there is rich historical data that could potentially be used to augment the control arms of adalimumab in future trials. We consider a new such trial in this paper. The trial objective is to develop a next-generation RA treatment that can outperform adalimumab. Due to confidentiality, the new treatment's information is not discussed. Our goal is to develop novel statistical models that borrow information from historical trial data involving adalimumab. In our effort to develop the models for the new trial, a collection of 11 historical trials using adalimumab is assembled. The rich information in this collection allows us to demonstrate the potential of the proposed model and design to increase the efficiency of the new trial. See the upcoming case study for more detail. 

From a regulatory perspective, historical controls have been permitted in \textit{confirmatory} trials primarily in rare and pediatric diseases as well as in devices \citep{ghadessi_roadmap_2020}. However, the regulatory threshold for their use is lower in non-confirmatory settings where there is less demand for conservative Type I error guarantees \citep{us_food_and_drug_administration_submitting_2019}. In order to use historical controls effectively, statistical models are critical due to the challenge in reconciling historical data and concurrent control data. Ideally, historical data that are more ``similar'' should be borrowed from more to aid statistical inference. The main questions are how to measure the similarity and how to ``borrow'' based on this measure.

Popular statistical methods for leveraging historical controls include propensity score approaches, which typically match or weight historical and current patients based on covariates \citep{lim_minimizing_2018, lin_propensity_2018, chen_propensity_2022}, as well as Bayesian modeling strategies including meta-analytic priors such as MAP and RMAP \citep{neuenschwander_summarizing_2010, schmidli_robust_2014}, power priors \citep{chen_power_2000}, commensurate priors \citep{hobbs_commensurate_2012},  multisource exchangeability models \citep{kaizer_bayesian_2018}, and LEAP \citep{alt_leap_2024}, which tend to borrow based primarily on similarities in response rates between the historical and new data. A barrier to using propensity score methods is that they require rich patient-level data, which is often unavailable (e.g. when such data are owned by a competing developer), and that researchers must both have \textit{and} select a sufficient set of covariates to control bias. While the Bayesian methods are attractive strategies to dynamically adapt the degree of historical borrowing, they typically assume a single mechanism of borrowing: \textit{either} that the historical and new data have similar unadjusted response rates, \textit{or} that they have similar response-covariate regression functions. As we show, this can lead to a loss of robustness or efficiency in scenarios violating the assumed borrowing mechanism. 

We propose a new model called SPx, standing for ``synthetic prior with covariates,'' to sharpen inferences about a new trial's control group response rate by borrowing from historical data. The main idea of SPx is to use a Bayesian expert system \citep{spiegelhalter_bayesian_1993} to merge different mechanisms of information sharing between the historical and new trial data. This strategy allows the model to efficiently tailor its inferences to different settings where distinct mechanisms of historical borrowing (or none) are appropriate. The model can be used simply for the analysis of a completed trial, but we also discuss how it can be embedded in an adaptive trial design to reduce the needed control arm sample size.

Similar to popular methods like MAP and RMAP \citep{neuenschwander_summarizing_2010, schmidli_robust_2014}, SPx requires only \textit{summary statistics} from the historical trials, not patient-level data. Such trial-level information is routinely reported in publications and press releases, and includes sample size, response rate, eligibility criteria, and average patient demographics and pre-trial clinical measures. This means that researchers using SPx may potentially draw from many more historical trials than if they were to use methods that require patient-level data. While patient-level are ideal to use for historical borrowing, they are often unavailable due to ethical and confidentiality reasons. Nevertheless, when patient data are available, the SPx model can be modified easily to accommodate the new data structure. We discuss this point briefly at the end of the paper. 

The paper proceeds as follows. In Section \ref{section:spx_model} we introduce the specific statistical setting, related methods, and the SPx model. Section \ref{section:spx_design} describes a two-stage adaptive design that leverages SPx to reduce control group sizes. In Section \ref{section:spx_sim_study} we discuss results from an extensive simulation study that benchmarks the method's performance and sheds light on its approach to dynamic borrowing. In Section \ref{section:spx_case_study} we apply the SPx approach to the design and analysis of the trial in rheumatoid arthritis, and in Section \ref{section:spx_discussion} we conclude.

\section{The SPx Model} \label{section:spx_model}

\subsection{Basic Data Setting and Related Models}

\textit{Data structure.} Following \cite{schmidli_robust_2014}, we consider  summary statistics of patient-level data in historical clinical trials. Specifically, denote the data by $(y_h, n_h, \mathbf{x}_h)$ for trials $h = 1, \dots, H, H+1$. Trials $1, \dots, H$ are the historical trials and  trial $H+1$ is the new trial, $y_h$ is the number of responders in trial $h$'s control group (from a binary endpoint), $n_h$ is the number of control patients in trial $h$, and $\mathbf{x}_h$ is a $(p+1)$-dimensional vector containing $p$ group-level covariates for trial $h$'s control group as well as an intercept. The covariates in $\bm{x_h}$ can include both basic characteristics of trial $h$ (e.g. eligibility criteria) and group-level summaries of patient-level covariates in trial $h$ (e.g. the mean age).

The  sampling model  of the historical and new trial control data is simply $$ y_h | \mathbf{x}_h, \psi_h \overset{iid}{\sim} Bin(n_h, \psi_h), \hspace{1em} h = 1, \dots, H, H+1, $$ where  $Bin$ denotes a binomial distribution, and $\psi_h$, potentially a function of covariates $\mathbf{x}_h$, is the true response rate of the control arm from trial $h$. While we say ``control'' arm here, in the historical trials this arm may have been a ``treatment'' arm that is now the standard of care. 

The key modeling questions are how to specify a joint prior on the true response rates  $\psi_h$ across the historical and new trials and how to take advantage of the covariates. The prior that we propose, SPx, extends and combines ideas from two popular strategies in the literature, which we now review. 

\textit{Meta-analytic priors.} The MAP, or meta-analytic predictive prior \citep{neuenschwander_summarizing_2010} is exchangeable, and models the logit response rates $\theta_h := \text{logit}(\psi_h)$ of all the trials (historical and new) with a common normal prior distribution: 
$$ \theta_1, \dots, \theta_H, \theta_{H+1} | \mu, \tau^2 \overset{iid}{\sim} N(\mu, \tau^2). $$
The prior mean $\mu$ and variance $\tau^2$ recieve hyperpriors to complete the hierarchical model specification. Although \cite{neuenschwander_summarizing_2010} emphasize the predictive interpretation of the MAP prior, here we focus on its hierarchical interpretation to illustrate how it induces borrowing through shrinkage. In MAP the variance parameter $\tau^2$ controls the degree of historical borrowing: if $\tau^2$ is small then the posterior will shrink $\theta_{H+1}$ strongly towards the historical data, and if $\tau^2$ is large then historical borrowing will be curtailed. Thus the hyperprior for $\tau^2$ is crucial to the performance of the MAP method, and in the original work \cite{neuenschwander_summarizing_2010} recommended sensitivity analysis for this part of the prior. In contrast, MAP's performance does not depend as heavily on the hyperprior for $\mu$ as long as it is not unreasonably concentrated. Often a noninformative uniform prior is used for $\mu$. The MAP model's popularity stems from its simple, familiar random effects model and good performance when the historical control rates are largely similar to the new trial's control rate. However, when the historical data are misleading the MAP approach often continues to borrow too heavily and thus lacks robustness.

Due to the MAP approach's strong reliance on the historical data's relevance to avoid bias and inflated Type I error, a Robust MAP (RMAP) model was also proposed \citep{schmidli_robust_2014}. This extension is a combination of the MAP prior and an independent prior that involves no borrowing from historical data, given by
\begin{align*}
    \theta_1, \dots, \theta_H | \mu, \tau^2 &\overset{iid}{\sim} N(\mu, \tau^2), \text{\hspace{1em} and} \\
    \theta_{H+1} | \mu, \tau^2 &\overset{ind.}{\sim} \pi N(\mu, \tau^2) + (1 - \pi) Logistic(0, 1).
\end{align*}
The componenent $Logistic(0,1)$ gives rise to robustness as it does not allow shrinkage of $\theta_{H+1}.$
Back on the probability scale, the $Logistic(0, 1)$ component is equivalent to a $Unif(0, 1)$ prior. In contrast to MAP, in RMAP the hyperprior on $\tau^2$ is less essential for limiting bias because of the inclusion of the independent component. However, the prior weight $\pi$ on the historical borrowing component is pre-specified and must be tuned through simulation to reliably control bias and Type I error. Based on that work $\pi = 0.5$ may be a reasonable default value in some settings.

\textit{Commensurate priors.} Another approach to historical borrowing is the commensurate prior \citep{hobbs_commensurate_2012, hobbs_adaptive_2013}, which explicitly models the difference between the historical and new trials \citep{pocock_combination_1976}. In our setting it first assumes that all historical rates are equal and then specifies the joint prior on the historical and new rates through a marginal prior on the historical rate and a ``commensurate'' prior on the new rate given the historical one:
\begin{align*}
    \theta_1 =  \dots = \theta_H = \theta | \mu, \tau^2 &\sim N(\mu, \tau^2), \text{\hspace{1em} and} \\
    \theta_{H+1} | \theta &\sim N(\theta, \sigma^2).
\end{align*}
Thus the (logit) difference between the new trial's rate and the historical rate is modeled as $N(0, \sigma^2)$, where the variance $\sigma^2$ is the key parameter controlling the amount of historical borrowing. Despite that assuming homogeneity of the response rates in the historical trials may be an oversimplification, this conditional specification provides a different mechanism to control borrowing than in the meta-analytic approaches. In past work \citep{hobbs_commensurate_2012} commensurate priors have been designed to work with patient-level data and covariates.

\subsection{The SPx Prior}
In this section we first describe the proposed prior intuitively, and then discuss the full details. SPx uses Bayesian model averaging (BMA) to combine model elements that allow different mechanisms of information sharing between historical trials and the currend trial. In particular, SPx first models the historical rates as conditionally exchangeable given covariates. Then the prior for the new rate, $\theta_{H+1}$, is a combination of three alternative submodels, or ``experts'':
\begin{enumerate}[Expert 1., leftmargin=2cm]
    \item a commensurate model, which directly assumes that the new rate is close to the historical rates (i.e. $\theta_{H+1}$ is centered on a weighted average of $\theta_1, \dots, \theta_H$);
    \vspace{-0.6em}
    
    \item a regression model, which predicts the new trial's rate based on its covariates $\mathbf{x}_{H+1}$ and the same covariate-response relationship present in the historical data (i.e. $\theta_{H+1}$ is centered on a regression prediction); and
    \vspace{-0.6em}
    
    \item an independent no-borrowing model, which assumes that the new rate is completely unrelated to the historical rates (i.e. $\theta_{H+1}$ is independent of $\theta_1, \dots, \theta_H$).
\end{enumerate}
A key distinction with RMAP is that SPx uses submodels that are well-separated in the sense that they put most of their prior mass on separate regions of the parameter space for $\left( \theta_1, \dots, \theta_{H+1} \right)$. This feature is motivated by the literature on nonlocal priors \citep{johnson_use_2010}. As we later discuss, it leads to posterior inferences about the mechanism and degree of historical borrowing that adapt more quickly to signals in the data. We now describe each component of the SPx prior in detail.

\paragraph{Priors for $\theta_h$ $(h=1, \dots, H)$ and $\theta_{H+1}$.} We assume that (i) the historical response rate $\theta_h$ for trial $h = 1, \dots, H$ can be modeled by a regression on covariates, and (ii) the new response rate $\theta_{H+1}$'s prior is a combination of three experts each using a different borrowing mechanism: direct historical borrowing, regression (on covariates), and no borrowing. In particular,
\begin{equation} \label{spx_prior_joint}
    \begin{aligned}
        \theta_h | \bm{\beta}, \tau^2, \mathbf{x}_h &\overset{ind.}{\sim} N(\bm{\beta}^T \mathbf{x}_h, \tau^2), \hspace{1em}  h = 1, \dots, H  \\
        \theta_{H+1} &= m_{hist} \underbrace{\theta_{hist}}_{\text{Expert 1}} + m_{reg} \underbrace{\theta_{reg}}_{\text{Expert 2}} + m_{ind} \underbrace{\theta_{ind}}_{\text{Expert 3}}.
    \end{aligned}
\end{equation}
Here the expert system, or model averaging, prior is obtained by assuming the Categorical prior 
$(m_{hist}, m_{reg}, m_{ind}) \sim Cat(p_{hist}, p_{reg}, p_{ind})$. In other words, the $m_k$ are binary indicator variables satisfying $m_{hist} + m_{reg} + m_{ind} = 1$, and $\theta_{H+1}$ is drawn from model $k \in \{hist, \phantom, reg, \phantom, ind\}$ with prior probability $p_k$. For example, with prior probability $p_{ind}$ the new trial's logit response rate is $\theta_{H+1} = \theta_{ind}$, which is independent of (completely unrelated to) the historical data. Priors for each expert, i.e. $\theta_{hist},$ $\theta_{reg}, \text{ and } \theta_{ind}$, are discussed next.

\paragraph{Priors for $\{\theta_{hist}, \theta_{reg}, \theta_{ind}\}.$} Below we use the terms ``expert'' and ``submodel'' interchangeably. The three prior submodels for $\theta_{H+1}$ are the key construction in SPx:
\begin{align*}
    & \text{Expert 1 (direct historical):} &  \theta_{hist} | \mu_{hist}, \sigma^2 &\sim N(\mu_{hist}, \sigma^2), \quad \mu_{hist} := \sum_{h=1}^H w_h \theta_h; \\
    & \text{Expert 2 (regression):} &  \theta_{reg} | \bm{\beta}, \tau^2, \mathbf{x}_{H+1} &\sim N(\bm{\beta}^T \mathbf{x}_{H+1}, c \tau^2);  \\
    & \text{Expert 3 (no-borrowing):} &  \text{logit}^{-1} (\theta_{ind}) &\sim Beta(0.5, 0.5). \\
\end{align*}

\vspace{-2em}

The prior for $\theta_{hist}$, the direct historical borrowing expert, is similar to the commensurate prior but does not assume that the historical rates are all equal. Instead, it assumes that $\theta_{hist}$ is centered at a weighted average of the historical rates. We define the weights $w_h$ to sum to 1 and be proportional to a distance metric between the regression's predicted response rates for trials $h$ and $H+1$:
$$ w_h \propto \left( 0.5 \right)^{\frac{| \widetilde{\psi}_h - \widetilde{\psi}_{H+1} |}{0.05}}$$
with $\widetilde{\psi}_h := \text{logit}^{-1}(\bm{\beta}^T \mathbf{x}_h)$. This choice encourages more borrowing from historical trials with covariates closer to $\mathbf{x}_{H+1}$, to the extent that the covariates predict response rates. In particular, the weights are scaled so that historical trial $h$ will receive twice the weight of any trial $j$ for which its predicted response rate is 5\% points closer to the predicted response rate of the new trial (i.e. $| \widetilde{\psi}_j - \widetilde{\psi}_{H+1} | - | \widetilde{\psi}_h - \widetilde{\psi}_{H+1} | = 0.05$). We tested other functional forms for the weights and found they typically do not affect posterior inferences for $\hat{\theta}_{H+1}$ greatly. We note that marginally (integrating out just the $\bm{\beta}$ parameter, which has a prior centered on zero), the weights are constant and thus the prior $p \left( \theta_{hist} | \theta_1, \dots, \theta_H \right)$ is centered at the unweighted average $H^{-1} \sum_{h=1}^H \theta_h$. This is more similar to the commensurate prior, which assumes a \textit{constant} historical rate across all previous studies (see Section 2.1). While simpler priors may be reasonable in some cases, using a weighted average of potentially varying historical rates with potentially varying weights is a less restrictive modeling assumption.

The prior on $\theta_{reg}$, the regression expert, assumes that the new trial follows the same covariate-response relationship seen in the historical trials. This occurs because the same coefficient $\bm{\beta}$ is used in the regression for $\theta_{h}$ ($h = 1, \dots, H$) in \eqref{spx_prior_joint} and for $\theta_{reg}$. Another key point is that the scale depends on the same $\tau^2$ indicating how successful the regression is at predicting the historical response rates, but modified by a constant $c$ so that, after marginalizing out $\tau^2$ and $\sigma^2$, the prior variance of $\theta_{reg}$ is similar to the prior variance of $\theta_{hist}$ (as we discuss below along with the hyperpriors). At the point $\bm{\beta} = \bm{0}$ the regression expert is equivalent to the classic MAP model. For the regression expert we have assumed a \textit{linear} regression for the relationship between trial-level covariates and control response rates because the number of historical trials $H$ is typically not large enough to estimate regressions with more complex functional forms. The linear regression may still be useful as a first-order approximation to more complex relationships, and as we discuss in Section 4.4 the use of BMA gives SPx robustness when the linear regression is misspecified. 

Lastly, $\theta_{ind}$ is the independent or no-borrowing expert and is included for robustness. We chose the $Logistic(0, 1)$ prior for the independent model because it is equivalent to a flat $Uniform(0, 1)$ prior on the response probability $\psi_{H+1}$, reflecting the ``conservative'' analysis that an investigator excluding historical data might choose \citep{schmidli_robust_2014}. This prior differs only slightly from the $Beta(0.5, 0.5)$ prior used by Expert 3 in SPx, which is a Jeffreys prior and may be slightly more efficient \citep{robert_harold_2009}. As we discuss in Supplementary Section 4, these priors produce nearly identical posterior inferences. 

To summarize, the full hierarchical model so far is given by

\begin{equation} \label{spx_model_hierarchical}
    \begin{alignedat}{2}
        \textbf{Likelihood:} \hspace{1em} && y_h | \mathbf{x}_h, \psi_h &\overset{iid}{\sim} Bin(n_h, \psi_h), \hspace{3em} h = 1, \dots, H, H+1 \\
        \textbf{Prior:} \hspace{1em} && \theta_h &:= \text{logit}(\psi_h), \hspace{3em} h = 1, \dots, H, H+1 \\
        && \theta_h | \bm{\beta}, \tau^2, \mathbf{x}_h &\overset{ind.}{\sim} N(\bm{\beta}^T \mathbf{x}_h, \tau^2), \hspace{3em}  h = 1, \dots, H \\
        && \theta_{H+1} &= m_{hist} \theta_{hist} + m_{reg} \theta_{reg} + m_{ind} \theta_{ind}, \\
        && \theta_{hist} | \mu_{hist}, \sigma^2 &\sim N(\mu_{hist}, \sigma^2), \hspace{3em} \mu_{hist} := \sum_{h=1}^H w_h \theta_h,  \\
        && w_h &\propto \left( 0.5 \right)^{\frac{| \widetilde{\psi}_h - \widetilde{\psi}_{H+1} |}{0.05}}, \hspace{3em} \widetilde{\psi}_h := \text{logit}^{-1}(\bm{\beta}^T \mathbf{x}_h) \\
        && \theta_{reg} | \bm{\beta}, \tau^2, \mathbf{x}_{H+1} &\sim N(\bm{\beta}^T \mathbf{x}_{H+1}, c \tau^2), \\
        && \text{logit}^{-1} (\theta_{ind}) &\sim Beta(0.5, 0.5), \\
        && (m_{hist}, m_{reg}, m_{ind}) &\sim Cat(p_{hist}, p_{reg}, p_{ind}).
    \end{alignedat}
\end{equation}

\paragraph{Prior model probabilities and model hyperpriors.} The SPx model \eqref{spx_model_hierarchical} is completed by hyperpriors for several important parameters. The prior model probabilities $(p_{hist}, p_{reg}, p_{ind})$ and the hyperpriors for the variances $\sigma^2$ and $\tau^2$ (and to a lesser degree $\bm{\beta}$) are important because they impact the posterior weighting of the different borrowing mechanisms in SPx. To see how, note that the posterior for $\psi_{H+1}$ is the average of the posteriors under each expert, weighting by the posterior probability that each is ``correct'': 
\begin{align} \label{spx_posterior}
    p \left( \psi_{H+1} | D \right) = \sum_{k \in \{hist, reg, ind \}} p(\psi_k | m_k = 1, D) \cdot p(m_k = 1 | D),
\end{align} 
where $D$ denotes the complete data $(y_h, n_h, \mathbf{x}_h)_{h=1, \dots, H+1}$. The posterior weights of each submodel $k \in \{hist, reg, ind \}$ may be written as 
\begin{equation} \label{spx_weights}
    p^*_k := p(m_k = 1 | D) = \frac{p(D | m_k = 1) \cdot p_k}{\sum_{j \in \{hist, reg, ind \}} p(D | m_j = 1) \cdot p_j},
\end{equation}
where $p(D | m_k = 1) = \int_{[0,1]^{H+1}} p(D | \psi_1, \dots, \psi_{H+1}) \cdot p(\psi_1, \dots, \psi_{H+1} | m_k = 1) \hspace{3pt} \mathrm{d}^{H+1} (\psi_1, \dots, \psi_{H+1})$ is the marginal likelihood, or evidence, of expert $k$. It is important to recognize that each expert's hyperprior affects their marginal prior on the control response rates, $p(\psi_1, \dots, \psi_{H+1} | m_k = 1)$. This in turn can greatly influence the expert's marginal likelihood and posterior model probability, changing the behavior of SPx overall. 

In light of this we set the hyperpriors with the goal of making SPx adaptively either (i) allow fairly aggressive historical borrowing or (ii) quickly transition to little historical borrowing, depending on the similarity between the current and historical trial data. To do so we both 
\begin{enumerate}[(a)]
    \vspace{-1em}
    \item make the priors in the $hist$ and $reg$ submodels relatively strongly concentrated (i.e. high prior probability of small variances $\sigma^2$ and $\tau^2$), and
    
    \item give relatively high prior weight to the $ind$ (no-borrowing) submodel (i.e. $p_{ind} > p_{hist}, p_{reg}$).
\end{enumerate}
In particular, we take 
\begin{align*}
    (p_{hist}, p_{reg}, p_{ind}) &= \left( \frac{1}{8}, \frac{1}{8}, \frac{3}{4} \right), \\
    \sigma &\sim TCauchy(0, 0.02, (0, \infty)), \\
    \tau &\sim TCauchy(0, 2.5, (0, \infty)),
\end{align*}
where $TCauchy(m, s, I)$ represents a Cauchy distribution with location $m$ and scale $s$ that has been truncated to the interval $I$. Here we use equal prior weights $p_{hist}$ and $p_{reg}$; if an analyst has an \textit{a-priori} belief that one of these submodels should be trusted more, the prior weights can be made unequal. While $\tau$ has a larger scale than $\sigma$, recall that the (conditional) prior variance of the regression submodel is $c \tau^2 = \nicefrac{1}{25} \cdot \tau^2$, not $\tau^2$, so the regression submodel makes similarly strong prior predictions as the direct historical borrowing one. The prior on the regression coefficients, which does not unduly affect inference, is
$$ \beta_i \overset{iid}{\sim} Cauchy(0, 2.5), \hspace{1em} i = 0, 1, \dots, p, $$
following from \cite{gelman_weakly_2008}.  

This choice of prior underlies SPx's novel strategy to dynamically determine how much historical borrowing is appropriate. Combining (a) and (b) makes SPx more robust to irrelevant historical data as needed in equation \eqref{spx_posterior} \textit{primarily} by increasing $p^*_{ind}$, the posterior probability of the no-borrowing submodel, and only to a much lesser degree by making the posterior inferences $p(\theta_k | m_k = 1, D)$ of the borrowing submodels more conservative. In further detail, (a) makes the borrowing submodels' marginal likelihoods decrease more quickly as disagreement between the new and historical data grows, and (b) accounts for the fact that the no-borrowing submodel makes less confident predictions overall and thus may have a relatively lower marginal likelihood even when it should be favored. While similar robustness might be sought by giving the borrowing submodels more diffuse priors, this can come at the cost of weaker borrowing when the historical data are actually relevant. We discuss these points further in the supporting information.

Effectively, through its well-separated BMA formulation, SPx allows the posterior for $\theta_{H+1}$ to make largely independent decisions on 1) \textit{whether} we should borrow from the historical data at all and (2) \textit{how strongly} we should borrow from those data, if we decide to. Standard commensurate prior models do not make this distinction since they have a unimodal prior on $\sigma^2$. RMAP does make this distinction, but only weakly. This is because it uses diffuse priors for $\tau^2$, meaning that the MAP component will still borrow somewhat conservatively even when the historical data are trustworthy; see, e.g. the credible interval widths in Table \ref{tab:spx_main}. Although our approach may not be the only strategy to achieve refined dynamic borrowing, we find it to be a useful device.

\begin{figure}[ht]
    \centering
    \includegraphics[width=\textwidth]{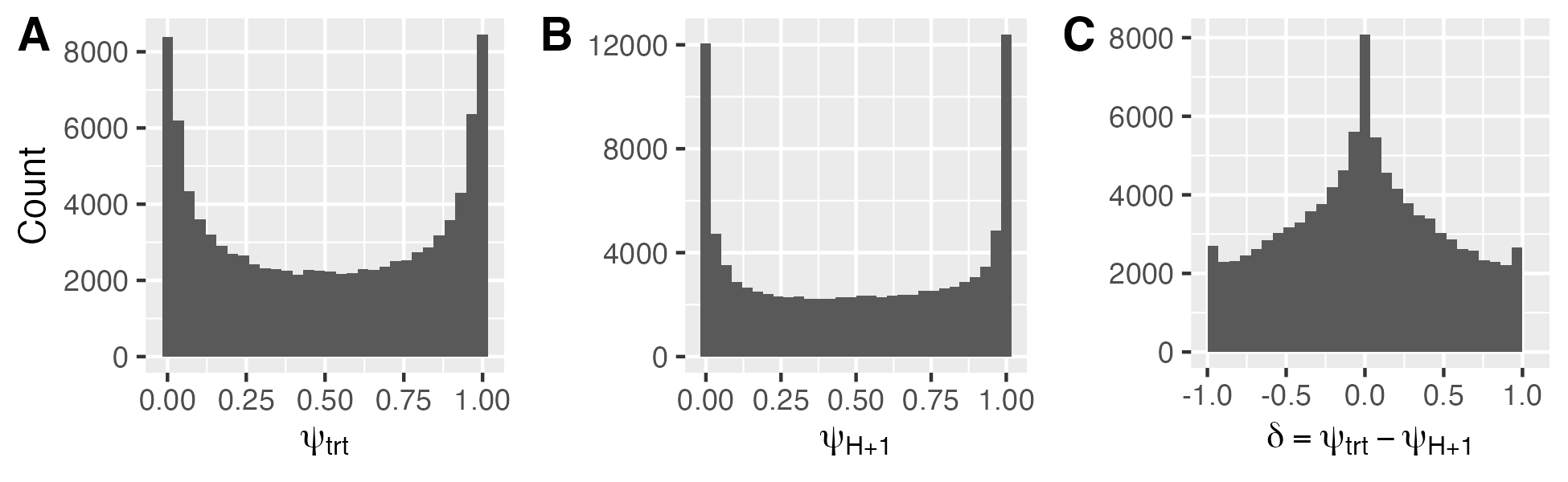}
    \caption{Histograms showing the marginal prior distributions for $\psi_{trt}$ (A), $\psi_{H+1}$ (B), and $\delta = \psi_{trt} - \psi_{H+1}$ (C) in the SPx model, based on 10,000 random draws. Priors shown are based on the historical and new trial covariate values used in simulation Scenario 1 (see Section 4).}
    \label{delta_fig}
\end{figure}

\textbf{Treatment effect estimation.}
To make inferences for the treatment effect we model the treatment group data $(y_{trt}, n_{trt})$ as independent from the historical and control data:
\begin{align*}
    y_{trt} | \psi_{trt} &\sim Bin(n_{trt}, \psi_{trt}) \\
    \psi_{trt} &\sim Beta(0.5, 0.5).
\end{align*}
Combined with the SPx prior for the new trial's control rate this induces a prior on the treatment effect, which we define as the difference $\delta := \psi_{trt} - \psi_{H+1}$ (although other comparisons, such as relative risk, could just as easily be used). Like with other popular methods \citep{schmidli_robust_2014}, in SPx the induced prior for $\delta$ does not have an analytic expression. By construction, the marginal prior for $\delta$ is symmetric with a mean of 0. This is due to the symmetry of the marginal priors for $\psi_{trt}$, $\text{logit}^{-1} \left( \theta_{hist} \right)$, $\text{logit}^{-1} \left( \theta_{reg} \right)$, and $\text{logit}^{-1} \left( \theta_{ind} \right)$, all of which are centered at 0.5 (a 50\% response rate). We illustrate the prior for $\delta$ in Figure \ref{delta_fig}.

\textbf{Computation.} Posterior computation for SPx can be done easily and efficiently using standard Bayesian MCMC tools. We implemented the model using JAGS (called from R) and it takes at most a few seconds to analyze a single data set on a personal computer. This produces draws from the posterior distribution of $\psi_{H+1}$, which can be combined directly with draws from the conjugate posterior for $\psi_{trt}$ to get draws from the posterior for the treatment effect $\delta$. For full details, see the JAGS and R code included in the supporting information.

\section{Adaptive Design Based on Posterior Inference} \label{section:spx_design}
We propose a two-stage adaptive design, largely following \cite{schmidli_robust_2014}, to reduce the control group size to the extent that reliable information can be gained through historical borrowing. Intuitively, in Stage 1 the new trial enrolls a fixed and prespecified number of control group patients. Then there is an interim check to determine the number of control patients to enroll in Stage 2. The interim check measures the degree of ``compatibility'' between the historical and new trial (Stage 1) data in terms of an effective sample size of control patients gained by borrowing from the historical data. If the new trial's stage one control data is deemed less compatible with the historical data, little borrowing will happen and the Stage 2 size will not be reduced much or at all. This limits the impact of prior assumptions about the relevance of the historical data on the overall trial size. Because we model the treatment group data independently from the control group data (having no shared parameters), the treatment group patients may be enrolled without reference to the control patients or the stages of the adaptive design (i.e. by simply changing the randomization probability after the interim check).

To define the adaptive design, we introduce notation for the data from various sources and stages of the trial. Let the historical data be denoted as $D_{hist} = (y_h, \bm{x}_h, n_h)_{h = 1, \dots, H}$, let the new Stage 1 data (control and treatment) be denoted as $D_1 = \left(y_{H+1, 1}, \phantom, \bm{x}_{H+1, 1}, \phantom, n_{H+1, 1}^c \right) \phantom, \cup \phantom, \left(y_{trt, 1}, \phantom, \bm{x}_{trt, 1}, \phantom, n_{trt, 1}^c \right)$, and let the new Stage 2 data (control and treatment, not including patients from Stage 1) be denoted as $D_2 = \left(y_{H+1, 2}, \phantom, \bm{x}_{H+1, 2}, \phantom, n_{H+1, 2}^c \right) \phantom, \cup \phantom, \left(y_{trt, 2}, \phantom, \bm{x}_{trt, 2}, \phantom, n_{trt, 2}^c \right)$. The data we analyze at interim are $D_I = D_{hist} \cup D_1$ and the data we analyze at the final analysis are $D_F = D_{hist} \cup D_1 \cup D_2$.

Formally, the adaptive design proceeds as follows:
\begin{itemize}[wide=0.5\parindent, leftmargin = 0.5\parindent]
    \item[\textbf{Stage 1}:] Collect data $D_1$ on $n_{trt, 1}^c$ treatment patients and $n^c_{H+1,1}$ control patients.
    
    \item[\textbf{Interim analysis}:] \hfill \vspace{-1em}
    \begin{itemize}
        \item[(a)] Compute the interim SPx posterior of $\psi_{H+1}$ given the interim data $D_I = D_{hist} \cup D_1$, i.e. $p \left( \psi_{H+1} | D_I \right)$.

        \item[(b)] Find the \textit{effective sample size} of this posterior, $n^c_{H+1, \phantom, eff}$, via the moment matching method of Weber (2020).
    \end{itemize}
    \vspace{-0.75em}
    
    \item[\textbf{Stage 2}:] Collect data $D_2$ on $n_{trt, 2}^c$ treatment patients and $n^c_{H+1,2}$ control patients, where $n^c_{H+1,2} = n^c_{H+1,max} - n^c_{H+1, eff}$ but is truncated to the interval $[\gamma_{min} n^c_{H+1, \phantom, max}, \phantom, \gamma_{max} n^c_{H+1, \phantom, max}]$ for some $\gamma_{min} \in [0, 1]$, $\gamma_{max} \in [1, \infty)$ (e.g. $\gamma_{min} = 0.75$, $\gamma_{max} = 1.25$).

    \item[\textbf{Final analysis}:] Inference for the treatment effect $\delta = \psi_{trt} - \psi_{H+1}$ is based on the SPx posterior given the final data $D_F = D_{hist} \cup D_1 \cup D_2$, i.e. $p \left( \delta | D_F \right)$.
\end{itemize}

The second stage is designed so that the total sample size of the control group is not intolerably lower or higher than the target size $n^c_{H+1,max}$.

We emphasize that inferences for $\psi_{H+1}$ and $\delta$ (as well as all other parameters) follow standard Bayesian principles at both interim and final analysis. In particular, the posterior distribution used for the interim analysis is
\begin{equation} \label{interim_psi_posterior_text}
    \begin{split}
        p \left( \psi_{H+1} | D_I \right) &\propto p \left( D_I | \psi_{H+1} \right) p \left( \psi_{H+1} \right) \\
        &\propto \int p \left( D_I | \bm{\psi}, \phantom, \bm{\beta}, \phantom, \tau, \phantom, \sigma, \phantom, \bm{z} \right) p (\bm{\psi}, \phantom, \bm{\beta}, \phantom, \tau, \phantom, \sigma, \phantom, \bm{z}) d \psi_{H+1}^{-},
    \end{split}
\end{equation}
where $\bm{\psi} = \left( \psi_1, \dots, \psi_H, \phantom, \psi_{H+1}, \phantom, \psi_{trt} \right)$ and $\bm{z} = (m_{hist}, \phantom, m_{reg}, \phantom, m_{ind})$ and $d \psi_{H+1}^{-}$ is shorthand to indicate that we integrate over all parameters in SPx other than $\psi_{H+1}$. The posterior of $\psi_{H+1}$ at the final analysis is defined similarly, with $p(D_F | \psi_{H+1})$ replacing $p(D_I | \psi_{H+1})$: 
\begin{equation} \label{final_psi_posterior_text}
    \begin{split}
        p \left( \psi_{H+1} | D_F \right) &\propto p \left( D_I | \psi_{H+1} \right) p \left( \psi_{H+1} \right) \\
        &\propto \int p \left( D_F | \bm{\psi}, \phantom, \bm{\beta}, \phantom, \tau, \phantom, \sigma, \phantom, \bm{z} \right) p (\bm{\psi}, \phantom, \bm{\beta}, \phantom, \tau, \phantom, \sigma, \phantom, \bm{z}) d \psi_{H+1}^{-} \\
        &\propto p \left( D_2 | \psi_{H+1} \right) \cdot p \left(\psi_{H+1} | D_I \right).
    \end{split}
\end{equation}
The third line in \eqref{final_psi_posterior_text} expresses the final posterior as a standard Bayesian update of the interim analysis \citep{berry_bayesian_2010}, where the new data is $D_2$ and the ``prior'' is the interim posterior, $p \left(\psi_{H+1} | D_I \right)$. The posterior for $\delta$ at the final analysis is induced by the posterior for $\psi_{H+1}$ and the independent posterior for $\psi_{trt}$ (as discussed in Section 2). 

Crucial to this design is the posterior effective sample size of SPx at interim, $n^c_{H+1, \phantom, eff}$, which is intended to assess how much information the historical data contribute about the new trial's control rate and how many fewer patients the new control group needs in exchange for this additional information \citep{hobbs_adaptive_2013, schmidli_robust_2014}. We use a simple moment matching definition of effective sample size (Weber 2020), which finds the Beta distribution with the same mean and variance as the SPx posterior, takes the effective sample size of this Beta distribution (the sum of its parameters; \cite{morita_determining_2008}), and subtracts the current sample size at interim, $n^c_{H+1,1}$. We note that it is technically possible that $n^c_{H+1, eff} < n^c_{H+1,1}$, indicating that using the historical data reduce certainty about the new trial's control rate and the Stage 2 sample size should be slightly increased to compensate. In truth, defining an effective sample size that has desirable properties for complex non-conjugate models such as SPx is an open area of research \citep{morita_determining_2008, morita_prior_2012, neuenschwander_predictively_2020}. However, the sample size thresholds used in Stage 2 of the design somewhat limit the impact of the choice of definition. The simple definition we use has the benefit of producing more conservative (smaller) effective sample sizes for SPx than other definitions, which sometimes produce implausibly large effective sample sizes for the SPx model. Regardless of the specific definition used, by measuring the effective sample size during, and not before, the trial we can measure the extent to which the new data diverge from the historical data, potentially safeguarding against inappropriate borrowing from irrelevant historical data. 

At the end of the trial, the decision rule to detect a treatment effect may take a variety of forms. Depending on the disease and regulatory setting, interest may focus on detecting either positive or clinically significant effects. To detect positive effects, we may use the rule 
\begin{align} \label{decision_positive}
    P(\delta > 0 | D_F) &\geq 1 - q_{positive}
\end{align}
for the treatment effect $\delta := \psi_{trt} - \psi_{H+1}$ where $q_{positive} \in (0, 1)$ is the posterior Type I error threshold. Alternatively, to detect clinically significant effects, we may use the rule
\begin{align} \label{decision_clin_sig}
    P(\delta > \delta_0 | D_F) &\geq 1 - q_{clinical}
\end{align}
for some minimal threshold $\delta_0 > 0 $ and some $q_{clinical} \in (0, 1)$. In our simulation and case studies we use $q_{positive} = q_{clinical} = 0.05$ to target a 5\% level of Bayesian type I error. However, in settings where strict Frequentist type I error control is needed researchers may tune these thresholds by simulation, which is a common regulatory requirement for complex trials using adaptive designs and/or historical data (U.S. Food and Drug Administration 2019). 

\section{Simulation Study} \label{section:spx_sim_study}

How accurately does the SPx model estimate the new trial's control response rate, and does it perform respectably when the historical data are misleading and borrowing would be detrimental? Further, when used with an adaptive design does SPx successfully reduce the trial's control group size while maintaining power and Type I error? 

To answer these questions we simulated clinical trials from scenarios defined by two factors: (i) whether or not the historical control rates are misleading (i.e. on average notably different from the new control rate), and (ii) whether or not the group-level covariates are associated with response rates. This yields four basic scenarios: Scenario 1 [ideal], where historical control rates are not misleading and covariates are predictive; Scenario 2 [covr], where historical rates \textit{are} misleading but covariates are still predictive; Scenario 3 [hist],  where historical rates are not misleading but covariates are \textit{not} predictive; and Scenario 4 [worst], where not only are historical rates misleading but also covariates are not predictive. In all cases the historical trials were loosely based on the real historical trials we analyze in Section \ref{section:spx_case_study}.

The full details of data generation for all four scenarios are provided in the supporting information. For each scenario, we generated a historical data set consisting of 15 historical trials ranging from 40 to 200 control patients each. We fixed that single data set, and repeatedly generated data for a new trial, 1,000 times for each method. This reflects the type of Frequentist repeated sampling we expect drug developers and regulators to be concerned with, since at the point of trial planning or analysis it is reasonable to imagine replicating the new trial but not also all of the historical ones. We also varied the target maximum sample size for the new trial. In all scenarios we simulated treatment group data for the new trial independently from the control data, at rates higher by both 0 and 30 percentage points. 

Scenarios 1 and 2 use the same historical data set in which the covariates are predictive. Similarly, Scenarios 3 and 4 use a different version of this data set in which the covariate measurements have been permuted to be uncorrelated with response rates. In all cases the observed historical control rates ranged from roughly 12\% to 38\%, with an average of 23\%. In Scenarios 1 and 3, the new trial's true response rate is 20\%, near the middle of the historical range. In Scenarios 2 and 4 it is 45\%, outside of the historical range. This means that direct historical borrowing in Scenarios 2 and 4 will bias estimates of the new trial's control rate downwards and thus estimates of the treatment effect estimates upwards, increasing both the Type I error and power of effect testing. The opposite would happen if the historical control response rates were higher than the new trial's response rate. Which situation is more likely in practice depends on how a variety of factors such as the standard of care, patient lifestyles, and patient demographics have changed over time.

We compared several methods including SPx, RMAP \citep{schmidli_robust_2014}, and an independent model (Ind.) with no historical borrowing (i.e. $\theta_{H+1} \sim Logistic(0, 1)$). For SPx and RMAP we included both versions where the new trial's control group size was fixed and where the two-stage adaptive design described in Section \ref{section:spx_design} was used (with $n^c_{H+1,1} = n^c_{H+1,max}/2$, $\gamma_{max} = 1.25$, and $\gamma_{min} = 0.75$). The two versions gave a contrast on the potential gain in the adaptive design in reducing the control group sample size.  The value $n^c_{max}$ is fixed at 200 or 100, which are the maximum control sample size. For SPx our standard implementation used only 2 of 6 covariates associated with response rates to mimic imperfect knowledge or data collection, but we also include cases where SPx uses all 6 covariates to show the effectiveness of the method when the regression is strong and not misspecified. For RMAP we used a 50-50 prior mixture between the borrowing (MAP) component and the non-borrowing component, since this specification performed well in \cite{schmidli_robust_2014}. The 90-10 mixture they consider was too aggressive and extremely biased in many of our scenarios. We emphasize that like SPx, RMAP adaptively updates the posterior weights (away from the 50-50 weights in the prior) for the two components depending on their relative model fits. This is consequence of the standard BMA formulation, and can be seen in equations (7) and (8) of \cite{schmidli_robust_2014} (similar to our equation \eqref{spx_weights}). For each method we computed the posterior mean and 95\% credible interval for $\psi_{H+1}$, as well as the posterior probabilities of $\delta$ described in equations \eqref{decision_positive} and \eqref{decision_clin_sig}. Note that these estimates and tests are Bayesian, so the credible intervals are not expected to have exactly 95\% Frequentist coverage and the tests are not expected to have exactly 5\% type I error rates.

\subsection{Estimation Accuracy for the New Trial's Control Rate}

Table \ref{tab:spx_main} shows performance of the models and designs for the task of reliably and efficiently estimating the new trial's control response rate. Although the overall goal of the trial is treatment effect testing, performance of the modeling strategies can be understood with more nuance by first considering control response rate estimation.

The results in Scenario 1 reveal that SPx can perform very strongly when the historical controls are directly relevant (i.e. their average response rate is close to the new trial's rate) and the covariates are predictive, as shown by competitive control group sizes and drastically improved RMSE and interval width, compared to RMAP and the independent model. In Scenario 3, where the historical controls are still directly relevant but the covariates are no longer predictive, SPx performs comparably to RMAP. 

%TC:ignore
\begin{table}[tb]

\centering

\resizebox{\textwidth}{!}{
\begin{tabular}{@{}llllllllllllllll@{}}
\toprule
 & \multicolumn{7}{c}{$\mathbf{n^c_{max} = 200}$} &  & \multicolumn{7}{c}{$\mathbf{n^c_{max} = 100}$} \\
 & \multicolumn{2}{c}{SPx} &  & \multicolumn{2}{c}{RMAP} &  & \multicolumn{1}{c}{Ind.} &  & \multicolumn{2}{c}{SPx} &  & \multicolumn{2}{c}{RMAP} &  & \multicolumn{1}{c}{Ind.} \\
 & Fixed & Adapt. &  & Fixed & Adapt. &  & Fixed & \hspace{1.5em} & Fixed & Adapt. &  & Fixed & Adapt. &  & Fixed \\ \midrule
\textbf{Scenario 1} &  &  &  &  &  &  &  &  &  &  &  &  &  &  &  \\
Size & 200 & 160.3 &  & 200 & 166.7 &  & 200 &  & 100 & 80.5 &  & 100 & 78.1 &  & 100 \\
RMSE & 0.018 & 0.018 &  & 0.026 & 0.028 &  & 0.029 &  & 0.022 & 0.022 &  & 0.031 & 0.033 &  & 0.039 \\
Coverage & 98.8 & 98.9 &  & 95.0 & 96.4 &  & 93.9 &  & 99.3 & 99.2 &  & 97.6 & 97.8 &  & 95.8 \\
Width & 0.086 & 0.094 &  & 0.103 & 0.112 &  & 0.110 &  & 0.117 & 0.130 &  & 0.138 & 0.152 &  & 0.155 \\ \arrayrulecolor{black!50}\midrule
\textbf{Scenario 2} &  &  &  &  &  &  &  &  &  &  &  &  &  &  &  \\
Size & 200 & 186.9 &  & 200 & 208.5 &  & 200 &  & 100 & 93.7 &  & 100 & 104.1 &  & 100 \\
RMSE & 0.032 & 0.032 &  & 0.036 & 0.036 &  & 0.034 &  & 0.049 & 0.050 &  & 0.058 & 0.060 &  & 0.049 \\
Coverage & 96.0 & 96.0 &  & 94.4 & 94.3 &  & 94.9 &  & 94.6 & 94.9 &  & 91.0 & 90.9 &  & 95.7 \\
Width & 0.130 & 0.134 &  & 0.140 & 0.137 &  & 0.137 &  & 0.184 & 0.190 &  & 0.200 & 0.196 &  & 0.191 \\ \midrule
\textbf{Scenario 3} &  &  &  &  &  &  &  &  &  &  &  &  &  &  &  \\
Size & 200 & 168.0 &  & 200 & 167.2 &  & 200 &  & 100 & 82.3 &  & 100 & 78.7 &  & 100 \\
RMSE & 0.025 & 0.026 &  & 0.025 & 0.027 &  & 0.029 &  & 0.032 & 0.033 &  & 0.033 & 0.036 &  & 0.039 \\
Coverage & 93.3 & 94.9 &  & 94.7 & 96.1 &  & 93.9 &  & 95.7 & 95.3 &  & 96.2 & 96.4 &  & 95.8 \\
Width & 0.097 & 0.103 &  & 0.103 & 0.112 &  & 0.110 &  & 0.129 & 0.140 &  & 0.138 & 0.153 &  & 0.155 \\ \midrule
\textbf{Scenario 4} &  &  &  &  &  &  &  &  &  &  &  &  &  &  &  \\
Size & 200 & 204.7 &  & 200 & 209.1 &  & 200 &  & 100 & 104.5 &  & 100 & 104.4 &  & 100 \\
RMSE & 0.035 & 0.035 &  & 0.037 & 0.036 &  & 0.035 &  & 0.050 & 0.050 &  & 0.054 & 0.056 &  & 0.049 \\
Coverage & 95.2 & 95.5 &  & 94.6 & 94.3 &  & 94.9 &  & 95.8 & 95.2 &  & 93.4 & 94.0 &  & 95.7 \\
Width & 0.138 & 0.136 &  & 0.140 & 0.137 &  & 0.137 &  & 0.197 & 0.192 &  & 0.201 & 0.196 &  & 0.191 \\ \arrayrulecolor{black}\bottomrule
\end{tabular}
}

\caption[SPx Accuracy and Trial Size]{Control group size and Frequentist estimation accuracy for the new trial's control response rate, averaged over 1,000 simulated trials. Metrics are defined as follows: size is the mean control group size; RMSE is the root mean square error of the posterior mean of the control group rate; coverage is the proportion of trials in which the 95\% quantile-based posterior credible interval for the control group rate contains the true rate; width is the mean width of these credible intervals. Note that the credible intervals are Bayesian and not designed or calibrated to give exactly 95\% Frequentist coverage. For the Ind Fixed method, the theoretical operating characteristics are identical in (a) Scenarios 1 and 3 (for a given sample size) and in (b) Scenarios 2 and 4 (for a given sample size) so we averaged results in these pairs.}

\label{tab:spx_main}

\end{table}
%TC:endignore

Notably, SPx can still perform well when the historical control rates are misleading as long as group-level covariates are moderately predictive of the rates (Scenario 2); in this case SPx was still able to reduce the control group size while maintaining similar accuracy to the no-borrowing Ind. approach, whereas RMAP was not. 

Scenario 4, where the historical data are entirely misleading, is especially challenging when the new trial's control group size is smaller. In this case, models that borrow from historical data have weaker evidence of conflict between the new and historical trials. However, here SPx does relatively well, with its RMSE closer to that of the no-borrowing approach.

Unsurprisingly, the performance of SPx is less rosy in the challenging setting of Scenario 4 where the historical data are entirely misleading, though a silver lining is that its Frequentist coverage only degrades slightly. Its RMSE also slightly edges out that of RMAP, but the no-borrowing approach is clearly much preferred here. This highlights the point that if there is significant concern about the relevance of the historical data then priors should be made more conservative to protect against the greater likelihood of bias. The straightforward way to achieve this in SPx would be to change the prior submodel probabilities to further favor the independent component. We conducted a small sensitivity analysis to demonstrate this point (see supporting information). As one might predict, bias is mitigated in Scenarios 2 and 4 at the cost of smaller efficiency gains from borrowing in Scenarios 1 and 3.
    
\subsection{Power and Type I Error for the New Trial's Treatment Effect}

Results for testing treatment effects largely follow from those on control rate estimation. We report the Type I error rate of declaring a \textit{positive} treatment effect when the effect is zero, and the power to declare a \textit{clinically significant} effect of 20\% ($\delta = 0.2$ in equation \eqref{decision_clin_sig}) when the true effect is moderately large (i.e. the true $\delta = 0.3$). 

Table \ref{tab:spx_mainpower} examines the statistical power of the adaptive designs assuming 2 and 6 covariates were included in SPx, respectively. Type I error is relatively well controlled by all methods. Of course, a drug developer or regulator requiring that Frequentist Type I error be more strictly controlled can calibrate the model or decision rule by simulation under the specific scenarios they are concerned about. This is a reality of all Bayesian trial methods and many Frequentist ones as well \citep[e.g.][]{lewis_bayesian_2007}, a point we revisit in the Discussion. 

The power of SPx tends to be greater or similar to the power of RMAP. Compared to the no-borrowing approach, SPx has better (Scenario 1), similar (Scenario 3), or slightly lower (Scenario 2) power except in Scenario 4, all while substantially reducing the control group size when allowed. 
% Scenario 4 is where SPx has more substantial power loss compared to not borrowing since neither the historical data nor covariates are informative of the new trial's control rate. In reality, if there is a strong belief that historical data are not reliable and covariates are not informative, SPx should not be considered and more importantly, one might not want to try borrowing information from historical data in the first place. 

%TC:ignore
\begin{table}[hbt]

\centering

\resizebox{\textwidth}{!}{
\begin{tabular}{@{}lllllllllllllllll@{}}
\toprule
 &  & \multicolumn{7}{c}{$\mathbf{n^c_{max} = 200}$} &  & \multicolumn{7}{c}{$\mathbf{n^c_{max} = 100}$} \\
 &  & \multicolumn{2}{c}{SPx} &  & \multicolumn{2}{c}{RMAP} &  & \multicolumn{1}{c}{Ind} &  & \multicolumn{2}{c}{SPx} &  & \multicolumn{2}{c}{RMAP} &  & \multicolumn{1}{c}{Ind} \\
 &  & Fixed & Adapt. &  & Fixed & Adapt. &  & Fixed & \hspace{1.5em} & Fixed & Adapt. &  & Fixed & Adapt. &  & Fixed \\ \midrule
$\mathbf{\delta}$ & \textbf{Scenario 1} &  &  &  &  &  &  &  &  &  &  &  &  &  &  &  \\
 & Size & 200 & 160.3 &  & 200 & 166.7 &  & 200 &  & 100 & 80.5 &  & 100 & 78.1 &  & 100 \\
  0 & $P(\delta > 0)$ & 5.4 & 4.3 &  & 4.6 & 3.8 &  & 5.5 &  & 3.2 & 2.7 &  & 1.5 & 1.2 &  & 4.7 \\
0.3 & $P(\delta > 0.2)$ & 78.9 & 77.4 &  & 65.6 & 63.6 &  & 70.3 &  & 54.6 & 51.4 &  & 41.2 & 38.7 &  & 46.7 \\ \arrayrulecolor{black!50}\cmidrule(l){2-17} 
 & \textbf{Scenario 2} &  &  &  &  &  &  &  &  &  &  &  &  &  &  &  \\
 & Size & 200 & 186.9 &  & 200 & 208.5 &  & 200 &  & 100 & 93.7 &  & 100 & 104.1 &  & 100 \\
  0 & $P(\delta > 0)$ & 3.8 & 3.8 &  & 6.3 & 6.0 &  & 4.9 &  & 5.5 & 5.5 &  & 10.5 & 10.4 &  & 5.7 \\
0.3 & $P(\delta > 0.2)$ & 68.5 & 65.9 &  & 69.3 & 70.7 &  & 69.1 &  & 43.7 & 42.3 &  & 47.2 & 48.0 &  & 43.1 \\ \cmidrule(l){2-17} 
 & \textbf{Scenario 3} &  &  &  &  &  &  &  &  &  &  &  &  &  &  &  \\
&  Size & 200 & 168.0 &  & 200 & 167.2 &  & 200 &  & 100 & 82.3 &  & 100 & 78.7 &  & 100 \\
  0 & $P(\delta > 0)$ & 2.3 & 1.7 &  & 2.9 & 2.6 &  & 5.5 &  & 2.2 & 1.7 &  & 2.2 & 2.0 &  & 4.7 \\
0.3 & $P(\delta > 0.2)$ & 70.1 & 66.3 &  & 70.4 & 66.3 &  & 70.3 &  & 44.6 & 40.4 &  & 44.3 & 37.9 &  & 46.7 \\ \cmidrule(l){2-17} 
 & \textbf{Scenario 4} &  &  &  &  &  &  &  &  &  &  &  &  &  &  &  \\
 & Size & 200 & 204.7 &  & 200 & 209.1 &  & 200 &  & 100 & 104.5 &  & 100 & 104.4 &  & 100 \\
  0 & $P(\delta > 0)$ & 5.5 & 5.2 &  & 6.6 & 5.9 &  & 4.9 &  & 6.2 & 6.3 &  & 8.2 & 8.8 &  & 5.7 \\
0.3 & $P(\delta > 0.2)$ & 69.8 & 70.2 &  & 70.1 & 71.4 &  & 69.1 &  & 45.2 & 45.7 &  & 47.8 & 50.1 &  & 43.1 \\ \arrayrulecolor{black}\bottomrule
\end{tabular}
}

\caption[SPx Type I Error and Power for Treatment Effects]{Frequentist Type I error and power for the new trial's treatment effect, averaged over 1,000 simulated trials. For convenience, the mean control group size is repeated from Table 1. Type I error is shown in the $P(\delta > 0)$ rows, where each model declares a positive treatment effect when its posterior probability that $\delta > 0$ exceeds 95\%; this row shows the \% of simulated trials where $\delta = 0$ in which each method instead declares $\delta > 0$. Power is shown in the $P(\delta > 0.2)$ rows, where each model declares a clinically significant treatment effect when its posterior probability that $\delta > 0.2$ exceeds 95\%; this row shows the \% of simulated trials where $\delta = 0.3$ in which each method declares $\delta > 0.2$. For the Ind Fixed method, the theoretical operating characteristics are identical in (a) Scenarios 1 and 3 (for a given sample size) and in (b) Scenarios 2 and 4 (for a given sample size) so we averaged results in these pairs.}

\label{tab:spx_mainpower}

\end{table}

\subsection{SPx's Adaptive Weighting of Borrowing Mechanisms} 

\begin{figure}[ht]
    \centering
    
    \includegraphics[width = \textwidth]{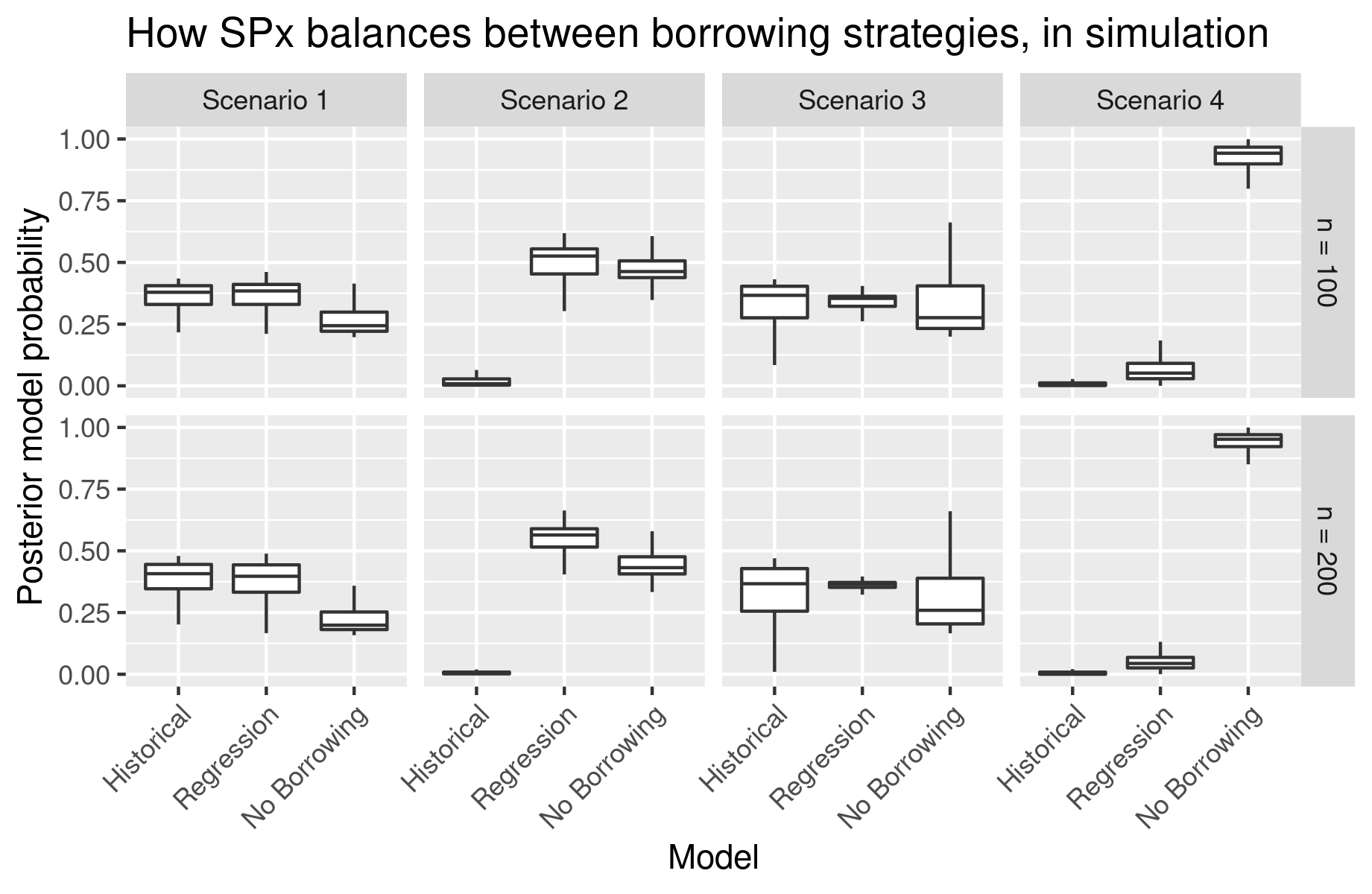}
    
    \caption[SPx BMA Behavior]{Boxplots of SPx's posterior submodel weights over 1,000 simulated trials in each scenario (columns) and trial size (rows). In these simulations the control group size of the new trial was fixed at 100 or 200 as shown.}
    
    \label{fig:spx_fig_1}
\end{figure}

To illustrate how SPx automatically adjusts the type of borrowing it performs depending on the historical and new trial data, Figure \ref{fig:spx_fig_1} plots the simulation distribution of SPx's posterior submodel weights $p^*_k$ (see equation \eqref{spx_posterior}) in Scenarios 1 through 4. When the historical data are entirely misleading (Scenario 4) SPx strongly favors its no-borrowing submodel. Otherwise SPx puts most posterior mass on its two borrowing submodels, appropriately favoring the regression submodel over the historical one when the historical rates are misleading but covariates are useful (Scenario 2). 

In the reversed setting where the historical rates are not misleading but covariates are not predictive (Scenario 3) SPx still gives moderate weight to the regression model because its inclusion of covariates adds some noise but not substantial bias. In particular, in Scenario 3 SPx correctly identifies that covariates are not predictive, meaning that the posterior distribution for the coefficient vector $\beta = (\beta_0, \beta_1, \dots, \beta_p)$ concentrates on a vector $(b_0, 0, \dots, 0)$ for some $b_0 \in \mathbb{R}$. The first element in this vector is the intercept in the regression function, representing the overall mean of the (logit) control response rates across studies. The remaining elements are the coefficients for the covariates, all of which are not associated with the outcome. In other words, in Scenario 3 the posterior of the regression submodel is similar to the simple MAP model shown in Section 2.1. This provides a similar inference on $\theta_{H+1}$ to the historical submodel, which also assumes that $\theta_{H+1}$ is close to a measure of the central tendency of $\theta_1, \dots, \theta_H$ (i.e. their weighted sample mean). Thus, in Scenario 3, it is sensible and expected that the regression and historical submodels would receive comparable weight.

It is also important to note that even when the historical data are relevant (Scenarios 1-3), in our simulations the independent no borrowing model still received non-trivial posterior weight (on average ~20\% - 50\% depending on the scenario). This is because SPx puts a conservative prior on the submodel weights, with 75\% prior probability going to the no-borrowing model. However, because the direct historical and regression submodels make aggressive predictions that heavily trust the historical data (see Section 3 and Supplementary Section S3), SPx still borrows significantly from the historical data in these scenarios. This is reflected in the increased efficiency of SPx's operating characteristics for Scenarios 1-3 in Tables \ref{tab:spx_main} and \ref{tab:spx_mainpower}. In this sense, the posterior submodel weights should not be interpreted by themselves as the sole summary of SPx's borrowing behavior.

\subsection{SPx's behavior when regression is biased for the new trial}

To illustrate how the regression and direct historical submodels contribute differently to SPx's Bayesian model averaging, we conducted additional simulations in which the regression submodel is correctly specified for the historical trial data but provides biased predictions for the new trial. These scenarios reflect situations where, for example, unmeasured confounding or systematic differences in covariate measurement make the historical regression function invalid in the new trial.

Scenarios 5 and 6 are modifications of Scenario 2. In Scenario 2, the regression model is correctly specified for both the historical and new trials. However, the new trial has extreme covariate values, so its true response rate is 44.5\% compared to an average of 23\% in the historical trials. In the new scenarios, we manually changed the new trial's true response rate while keeping the regression model, the historical trial data, and the new trial's covariates fixed. In Scenario 5, the new trial's response rate is 26\%, which is close to the historical average. In Scenario 6, it is 33\%, which is farther from the historical average but still well below the regression prediction of 44.5\%. These changes make the regression model correctly specified for the historical trials but misspecified for the new trial.

The results for Scenarios 5 and 6 (Supplementary Figure S2) show that SPx correctly and substantially down-weights the regression submodel, relying more heavily on the direct historical submodel (and, to a lesser extent, the independent submodel, especially in Scenario 6). As a result, SPx continues to provide accurate estimates of the new trial's control response rate (Supplementary Table S2) and accurate tests of its treatment effect (Supplementary Table S3). To summarize, in these scenarios the regression submodel is badly biased for the new trial. However, SPx limits its influence through Bayesian model averaging. SPx automatically identifies that the regression submodel is less predictive of the new trial's observed data (in terms of marginal likelihood, see equation \eqref{spx_weights}) and thus shifts its weight to better-performing alternatives, preserving the accuracy of the reported BMA inferences.

We note that for readers concerned about functional form assumptions (i.e. linearity) in the regression submodel, these simulation results offer some reassurance. What matters in SPx is not whether the regression submodel is strictly correct, but whether it provides accurate predictions compared to the other submodels, as measured by the marginal likelihoods. If a bad functional form assumption makes the regression submodel sufficiently inaccurate, it will be down-weighted in the SPx posterior in favor of better-performing submodels.

In settings where the new trial's covariates are clear outliers relative to the historical controls' covariates, extrapolation is unavoidable. In such cases, analysts should carefully consider whether the historical data should be used at all, regardless of the borrowing method. Even methods that do not explicitly model the response-covariate relationship can suffer from inappropriate extrapolation. For example, in propensity score methods, outlier covariate values in the new trial can violate the positivity (or common support) assumption \citep{westreich_invited_2010}. This substantially complicates the analysis, and the standard guidance is to remove units with extreme propensity scores --- effectively recommending that the historical and new trial data not be combined \citep{petersen_diagnosing_2012, hill_assessing_2013, zhou_propensity_2020}. SPx provides some automatic protection against such extrapolations because it will revert primarily to the independent submodel (see Supplementary Figure S2). However, if large covariate discrepancies between the new and historical trials are expected in advance or are observed in the data, we advise against using historical data for design and analysis, regardless of the method used.

\section{Rheumatoid Arthritis Case Study} \label{section:spx_case_study}

We discuss the application of SPx to the development of novel treatments for rheumatoid arthritis (RA). RA is an auto-immune disease that affects more than 1.3 million patients in the United States (Hunter et al. 2017), and it is predicted that its global burden will increase through 2030 \citep{cai_burden_2023}. The symptoms of this disease include pain and swelling joints in hands and feet as well as morning stiffness lasting longer than 30 minutes. Although there is no cure for RA, several treatments are available to slow down the disease progression and alleviate symptoms. Adalimumab is a well-established, standard biologic therapy in RA and has been approved for 20 years. As the patent of adalimumab is expiring globally and becoming a standard of care, novel therapies for RA must demonstrate superior treatment effects over adalimumab. We apply our SPx methodology in adalimumab trials conducted in the past two decades to explore possibilities for the design and analysis of a future trial in RA. Since adalimumab's own development was extensive, there are many past trials including an adalimumab arm that could be used as a rich source of historical data to potentially accelerate trials of novel RA drugs. In this process, SPx uses trial-level covariate information and a combination of historical borrowing mechanisms to guide the new trial. 

%TC:ignore
\begin{table}[hbt]

\centering
    
\addtolength{\tabcolsep}{5pt}  
    
\resizebox{\textwidth}{!}{
\begin{tabular}{@{}llllll@{}}
\toprule
\multicolumn{1}{c}{Trial} & \multicolumn{1}{c}{Reference} & \multicolumn{1}{c}{\begin{tabular}[c]{@{}c@{}}Previous\\ Treatment\end{tabular}} & \multicolumn{1}{c}{\begin{tabular}[c]{@{}c@{}}Average\\ Age\end{tabular}} & \multicolumn{1}{c}{Size} & \multicolumn{1}{c}{\begin{tabular}[c]{@{}c@{}}Response \\ Rate (\%)\end{tabular}} \\ \midrule
ALTARA & \cite{kennedy_efficacy_2014} & MTX & 48.8 & 43 & 39.5 \\
ARMADA & \cite{weinblatt_adalimumab_2003} & MTX & 56.0 & 62 & 21.0 \\
DE019 & \cite{keystone_radiographic_2004} & MTX & 56.1 & 200 & 24.0 \\
IM133-001 & \cite{weinblatt_efficacy_2015} & MTX & 51.4 & 61 & 39.3 \\
ORAL-Standard & \cite{van_vollenhoven_tofacitinib_2012} & MTX & 53.7 & 106 & 26.4 \\
RA-BEAM & \cite{taylor_baricitinib_2017} & MTX & 53.0 & 488 & 40.2 \\
STAR & \cite{furst_adalimumab_2003} & MTX & 55.8 & 315 & 29.5 \\
A3921035 & \cite{fleischmann_phase_2012} & none & 53.0 & 59 & 22.0 \\
CHANGE & \cite{miyasaka_clinical_2008} & none & 53.4 & 87 & 12.6 \\
DE007 & \cite{van_de_putte_efficacy_2003} & none & 50.2 & 70 & 10.0 \\
DE011 & \cite{van_de_putte_efficacy_2004} & none & 53.5 & 110 & 18.2 \\ \bottomrule
\end{tabular}
}
    
\caption[Adalimumab Trials]{Trials included in the adalimumab case study. Previous treatment, average age, size, and response rate refer to those of the control group in each trial. MTX is an abbreviation for methotrexate. The response rate is the proportion of patients experiencing a 20\% improvement in joint health at 12 or 13 weeks on the American College of Rheumatology criterion (ACR20).}

\label{tab:spx_case_study}
    
\addtolength{\tabcolsep}{-5pt}  
    
\end{table}
%TC:endignore

We have collected group-level data from 11 past adalimumab trials,  as shown in Table \ref{tab:spx_case_study}. We note that Lim et al. (2018) provided a framework of objectively selecting historical trials to avoid cherry picking, and their strategy could be used to expand the data set to include even more historical adalimumab trials. We illustrate the use of SPx by borrowing from the placebo arms of these trials, though in practice the same could be done using the adalimumab arms. The primary endpoint we use is the popular, regulatory-approved binary endpoint of ACR20, which is whether or not a patient has a 20\% or greater improvement in joint health on the American College of Rheumatology criterion (ACR20) 12 or 13 weeks after treatment. The first trial-level covariate we use is whether patients in the trial had previous or ongoing treatment with methotrexate (MTX), a common first line therapy for rheumatoid arthritis. Unsurprisingly, Table \ref{tab:spx_case_study} suggests that MTX use is a very strong predictor of ACR20 rates: the MTX trials have rates ranging from roughly 20-40\% while the no-MTX trials have rates ranging from roughly 10-20\%. This suggests that the regression strategy embedded in SPx may be useful despite the modest number of trials. We also include average patient age at baseline, which may be a proxy for disease progression or otherwise relate to ACR20 rates.

\begin{figure}[ht]
    
    \centering
    
    \includegraphics[width = \textwidth]{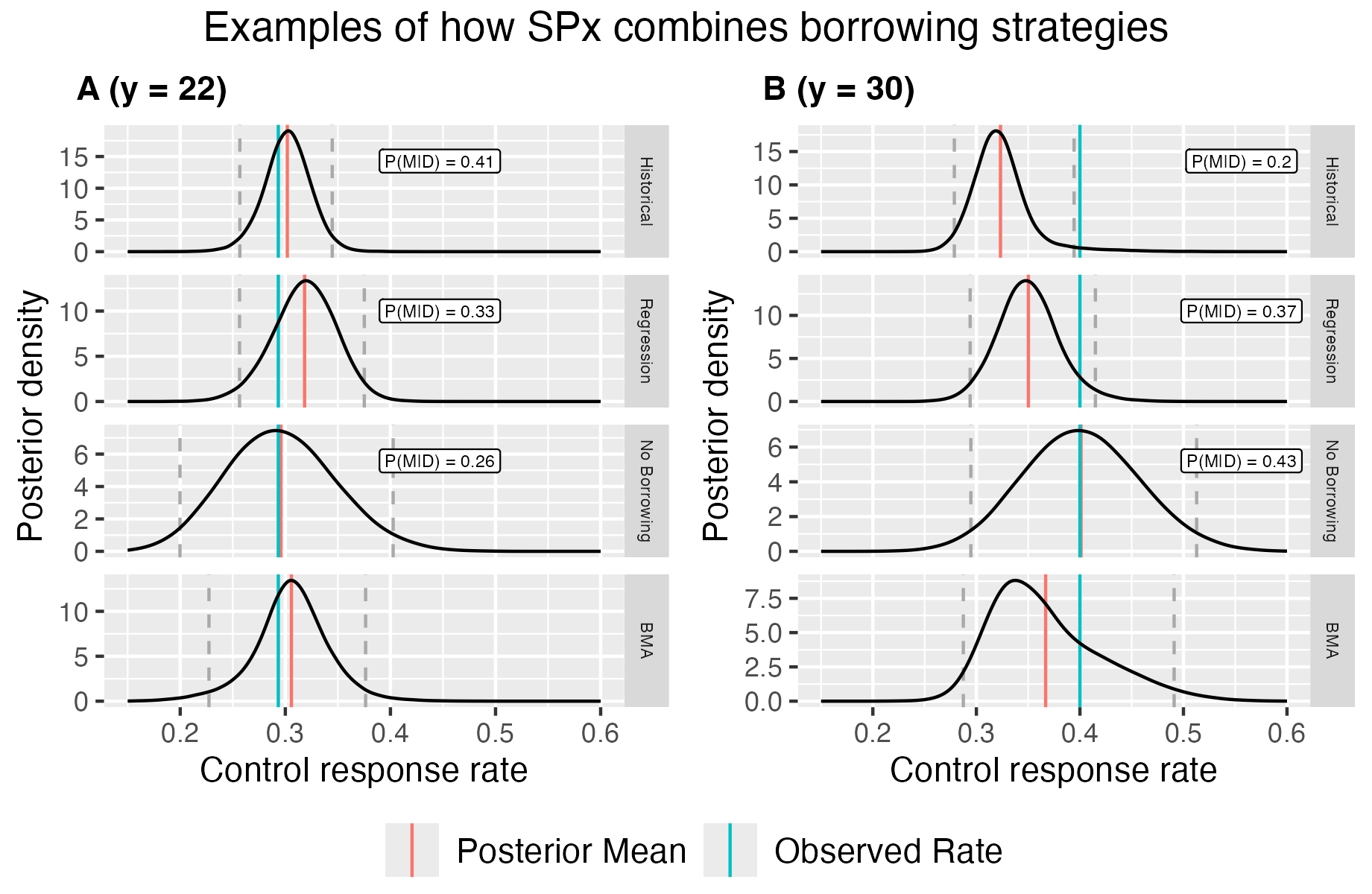}
    
    \caption[Example Adalimumab Posteriors]{Posterior distributions of the new trial's control response rate in two examples. In A, 22 of 75 control patients in the new trial are responders, and in B 30 of 75 are responders. The top 3 panels in each show the posterior distribution for each of SPx's submodel, along with the submodel's posterior model probability $P(M|D)$. The bottom panel shows the model averaged posterior, which is the SPx inference.}
    
    \label{fig:spx_fig_2}
\end{figure}

We now show how SPx would borrow information from this historical data set in the design and analysis of a new RA trial. We suppose that the new trial enrolls 75 control arm patients who have all had previous treatment with MTX and have an average age of 53. From Figure \ref{fig:spx_fig_2} we can see what the the BMA inference in SPx would be when the new control group's observed response is similar to (A, at 29.3\%) or far from (B, at 40\%) to the average historical response rates of 25.7\% overall and of 31.4\% among prior MTX trials. In panel A, SPx borrows heavily, giving 75\% of the posterior mass to to the relatively confident borrowing submodels. In panel B, SPx is more conservative; it gives more posterior mass to the no-borrowing submodel and its 95\% credible interval more or less reproduces that of the no-borrowing submodel. In this case it is suggested that there is enough conflict between the historical and new trial data that including the historical data in the analysis should \textit{not} increase our level of certainty about the new trial's control response rate.

\begin{figure}[ht]

    \centering
    
    \includegraphics[width = \textwidth]{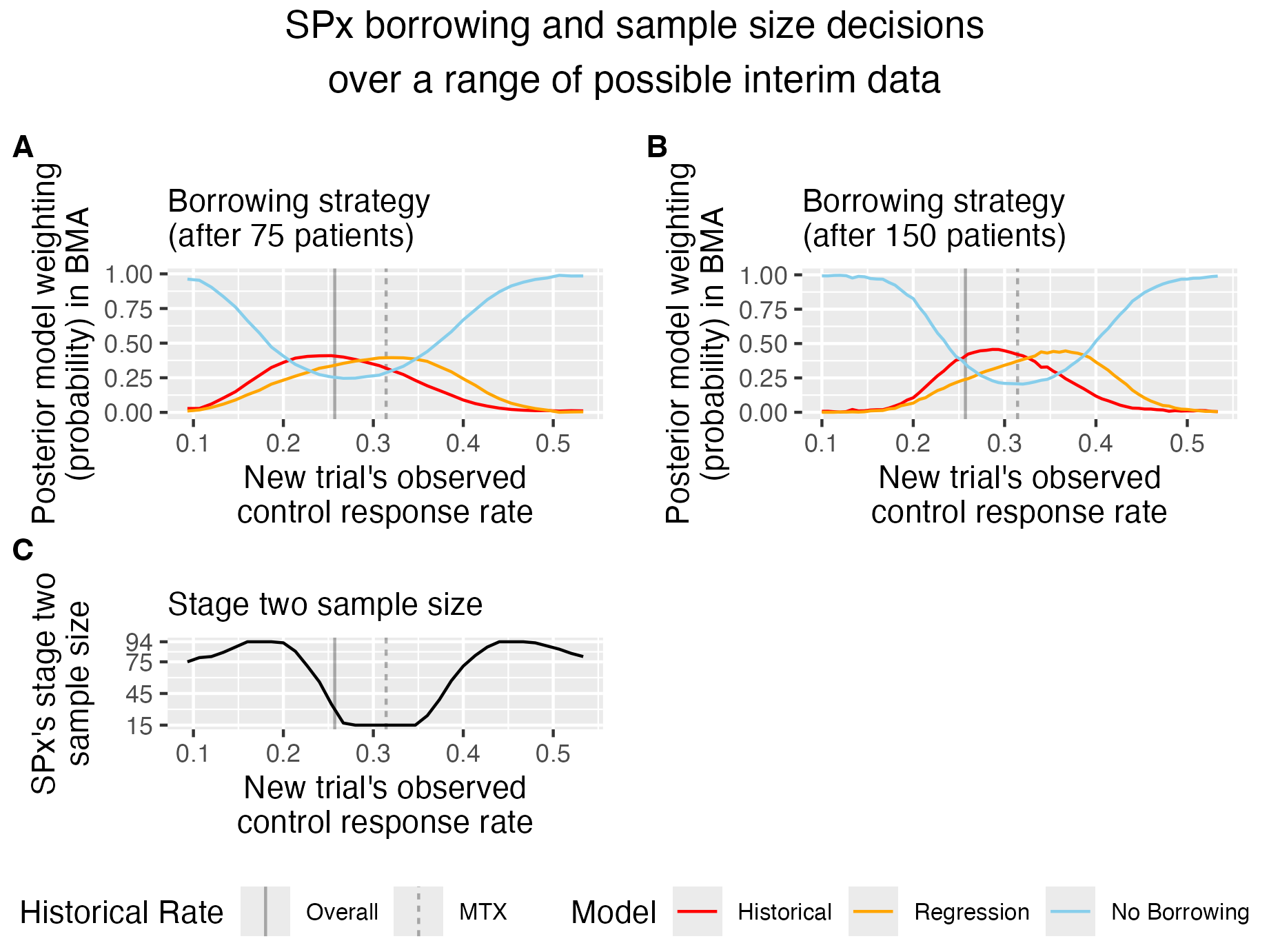}
    
    \caption[Range of Possible Adalimumab Posteriors]{A and B plot the range of posterior submodel probabilities in SPx as the new trial's observed response rate varies, after 75 and 150 patients respectively. C plots the range of interim sample size decisions as the new trial's observed response rate varies, after 75 patients. The vertical grey lines mark the sample mean of the historical response rates over all 11 trials (solid) and among just the prior MTX trials (dashed).}
    
    \label{fig:spx_fig_3}
\end{figure}

Figure \ref{fig:spx_fig_3} illustrates how SPx would adjudicate between its three borrowing mechanisms over a continuum of possible new trial data, with the observed response rate ranging from (roughly) 10\% to 50\%. Like in Figure \ref{fig:spx_fig_2}, the new trial's control arm has had previous MTX treatment and has an average age of 53. In panels A and B we can see that the degree of borrowing is largely controlled by how close the new trial's rate is to the observed overall and MTX-specific historical rates. The $hist$ and $reg$ submodels receive their maximum posterior weighting near these rates. In contrast, as the observed rate diverges from these historical rates the no-borrowing submodel quickly gains posterior weight, especially when the new trial's control group size is larger. Panel C shows how the SPx prior leads to different Stage 2 sample size determinations (based on the adaptive design described in Section 3) depending on the new trial’s observed response rate at interim. When the observed rate is close to the predictions of the $hist$ and $reg$ submodels, the Stage 2 sample size can be very small (compared to the target size of 75 with no borrowing). However, when the observed rate is \textit{far} from these historical predictions (i.e. $< 0.22$ or $> 0.4$ on the x-axis), the design can require a slightly \textit{larger} Stage 2 control group. In this case the modeling assumption to allow historical borrowing actually \textit{increases} uncertainty about the new trial's rate compared to not borrowing at all, so the design calls for collecting more data than otherwise planned. The profile of sample size decisions plotted in Panel C is influenced by the SPx model, and in particular the prior probability of the no-borrowing submodel, $p_{ind}$. If this profile were deemed unreasonable in practice, this may be a sign that the trialists should consider changing $p_{ind}$ to better reflect their beliefs. 

These results are valuable for the design and conduct of the new RA trial. Specifically, an interim analysis and sample size re-estimation based on the proposed SPx approach will potentially save resources and time for the new trial. Of course, the final outcome and efficiency gain depends on the interim data of the new trial. 

\section{Discussion} \label{section:spx_discussion}
The SPx method allows flexible borrowing from historical data using a novel Bayesian model averaging approach that balances between three mechanisms of borrowing: direct borrowing from the historical response rates, regression prediction using covariates, and no borrowing at all. The key methodological insight to improve the model averaging is to make the borrowing submodels give strong prior predictions that are well-separated from the no-borrowing submodel while giving high prior weight to the no-borrowing submodel. This strategy offers multiple avenues to not only take advantage of the historical data but also to avoid over-using it when the data suggest this may be unwise. If the group-level historical data are relevant to the new trial, then under a simple two-stage adaptive design SPx can considerably reduce its control group size, reducing the trial duration and cost.

Like the RMAP method, SPx uses Bayesian model averaging to adaptively balance between borrowing from the historical data and discounting it when appropriate. However, we highlight two important differences between SPx and RMAP. First, because SPx includes both a direct-borrowing component and a regression component, it will be able to robustly gain accuracy whenever \textit{either} (a) the new trial's control rate is similar to the historical control rates or (b) group-level covariates predict the response rates. In contrast, RMAP only gains accuracy in case (a). And second, SPx can be more robust than RMAP to different types of discrepancies between the historical and new control data, especially for smaller trials. See for example Table \ref{tab:spx_mainpower}, where SPx has better control of Type I error rate than RMAP in Scenarios 2 and 4 when $n^c_{max} = 100$. This is due to the careful construction of the SPx hyperprior (i.e. its borrowing submodels have concentrated, aggressive priors but the prior model weights favor the independent submodel), as we discuss in Sections 2.2 and 4.3.

Our simulation results are in line with the intuitive appeal of the SPx approach, and its performance is strong except when its prior assumptions are badly violated and the historical data are by no means useful. Aside from this exception, it produces more or similarly accurate estimates of the new trial's control group rate and treatment effect while substantially reducing the control group size. We recommend that trialists concerned with dramatically unfavorable scenarios include these in their design simulations and tune the prior or decision rules accordingly as is standard practice.

More generally, trialists wanting to prospectively calibrate the borrowing/no-borrowing tradeoff (i.e. power/Type I error tradeoff) for a specific planned trial may find the best success in tuning one of several key parameters of the method. First, they may experiment with increasing or decreasing $p_{ind}$, the prior probability of the no-borrowing submodel, as we do in Table S1. They may also consider making $p_{hist}$ and $p_{reg}$ unequal if one of these submodels is less suited for scenarios they are concerned about. Alternatively, they may experiment with changing the decision rule threshold $q_{positive}$ or $q_{clinical}$ for the trial's final treatment effect inference in equations (2) or (3), which would not require new simulation for each value if MCMC output has been saved. Both reducing $p_{ind}$ and $q_{clinical}$ or $q_{positive}$ would reduce Type I error at the cost of also reducing power. Further, it may be desirable to tune the values $n^c_{max}$, $\gamma_{max}$, and $\gamma_{min}$ in the adaptive design in order to achieve specific practical constraints or operating characteristics. 

In general, a variety of factors impact whether a trial setting is likely to benefit and safely reduce control group size by using SPx. The greatest benefit will come when the historical and new trials share data on group-level covariates that are strongly predictive of response rates and when there are enough historical trials to estimate this regression relationship with reasonable accuracy. This may include heavily studied disease areas and drugs (e.g. pembrolizumab) or those where treatments and outcomes have been slow to change (e.g. newly diagnosed glioblastoma). 

Because the method only requires group-level data, it may be possible to include trials where the patient-level data would not be available due to privacy or intellectual property concerns. The group-level covariates might ideally be believed to be strong predictors based on solid theoretical or past empirical evidence, though in settings with enough historical trials it may be possible to incorporate higher-dimensional covariates through the use of sparsity-inducing priors on the regression coefficients. We note that because we use trial-level summary statistics as covariates, one might be concerned that variability (i.e. measurement error) in these statistics could affect the performance of SPx. However, SPx uses these covariates only to improve prediction of the new trial's control response rate via the linear predictor $\bm{\beta}^\top \bm{x}_{H+1}$. It is well-established that while measurement error in covariates can bias estimation of regression coefficients \citep{carroll_measurement_2006}, it does not generally reduce prediction accuracy (\cite{carroll_measurement_2006}, Chapter 2.6; \cite{khudyakov_impact_2015}). For this reason, we do not expect measurement error in the group-level covariates to substantially impair SPx's predictions. An important exception arises if the magnitude or structure of measurement error differs substantially between the historical and new trials (\cite{carroll_measurement_2006}, Chapter 2.6; \cite{luijken_impact_2019}), and this is an important area for future work. In this case, it would be necessary to explicitly model the measurement errors.

While not the focus of the present work, using the SPx modeling strategy with patient-level data and covariates would likely reduce trial sizes and increase accuracy even more substantially. The proposed SPx model could easily be modified to accommodate the patient-level data. For example, notation $y_{hj}$ and $x_{hj}$ would represent the response and covariates for patient $j$ from study $h$. Then patient-specific random effects $\psi_{hj}$ may be introduced and a prior model $f(\psi_{hj} | \psi_h)$ may be used to allow shrinkage of the random effects to study-specific effect $\psi_h$. The SPx model can then be used for $\psi_h$ to complete the model construction. We also note that with patient-level data it may be feasible for analysts to use more sophisticated approaches to control for covariates (compared to our linear regression of trial-level covariates over a moderate number $H+1$ of studies). For example, with patient-level data one might consider replacing SPx's regression submodel with a semiparametric model that allows flexible covariate-response relationships \citep{zhou_incorporating_2021}, a model that integrates propensity scores \citep{lin_propensity_2018, chen_propensity_2022}, or other semiparametric Bayesian strategies \citep{muller_product_2011}.

\parindent 0cm
\parskip 0ex

%TC:ignore

\section*{Acknowledgments}

\small
\setstretch{1.2}
\parskip 5pt

This manuscript was partly sponsored by AbbVie. AbbVie contributed to the design, research, and interpretation of data, writing, reviewing, and approved the content. Li Wang is an employee of AbbVie Inc. and may own AbbVie stock.

\begingroup
    \small
    \setlength{\bibsep}{5pt}
    \setstretch{1}
    \bibliography{SPx.bib} 
\endgroup

\normalsize
%TC:endignore

\section*{Supporting Information}

Supporting information referenced in Sections 2 and 4 are available with this paper as supplementary files. R and JAGS code to estimate the models and run simulations are also available at this location.

\end{document}